\documentclass[aps,pra,a4paper,amsfonts,amssymb,amsmath,reprint,twoside,
superscriptaddress,onecolumn,nofootinbib,longbibliography,floatfix]{revtex4-2}

\usepackage{amsmath,amssymb,amsthm,mathtools,bm}
\usepackage{graphicx,booktabs,array}
\usepackage[export]{adjustbox}
\usepackage{tikz}
\usetikzlibrary{calc,positioning,arrows.meta,decorations.pathreplacing}
\usepackage[colorlinks=true,urlcolor=blue,linkcolor=blue,citecolor=blue]{hyperref}
\usepackage[capitalize]{cleveref}
\usepackage{braket,microtype,xurl}

\newcommand{\ceil}[1]{\left\lceil #1\right\rceil}

\newcommand{\xor}{\mathbin{\oplus}}

\hypersetup{
 pdftitle={Halving the cost of QROM again via dense encoding},
 pdfauthor={Danial Motlagh},
 pdfsubject={Dense encoding of pairs of table entries in dirty-ancilla QROM}
}

\begin{document}
\title{Halving the cost of QROM again via dense encoding}
\author{Danial Motlagh}
\affiliation{Xanadu, Toronto, ON, M5G 2C8, Canada}

\begin{abstract}
Quantum read only memory (QROM) is a substantial source of non-Clifford
cost in fault-tolerant quantum algorithms. Borrowing dirty qubits can
reduce this cost, provided their unknown initial states are restored.
We consider the practically relevant qubit-constrained regime, where
limited workspace restricts the table-block size and table loading
dominates the cost. For $N$ entries of $b$ bits and approximately
$b\lambda$ dirty qubits, SelectSwap has leading Toffoli count
$2N/\lambda$. Recent work using sequential bit packets reduces this
to $(1+1/b)N/\lambda$. In this work we introduce a dense encoding
construction of QROM, where two bits of classical information are stored
in each dirty qubit instead of one by utilizing both the $Z$ and $X$ bases.
On its own, dense encoding provides a $2\times$ improvement over
SelectSwap by reducing the leading Toffoli count to $N/\lambda$.
Combining it with sequential bit packets gives
$\tfrac12(1+1/b)N/\lambda$, a further halving of the sequential
construction's leading cost, and an approximately $4\times$ improvement
over SelectSwap in the large-$N$ limit for practical values of $b$.
\end{abstract}
\maketitle
\setlength{\parskip}{\baselineskip}
\setlength{\parindent}{0pt}

\section{Introduction}
\label{sec:introduction}

Coherent lookup of classical data is an essential operation in fault-tolerant
quantum algorithms. Applications range from Hamiltonian simulation
\cite{Berry2019,VonBurg2021,Lee2021}, often built on quantum signal
processing~\cite{LowChuang2017,MotlaghWiebe2024}, to state preparation
\cite{Babbush2018,RupprechtWoelk2026} and unitary synthesis
\cite{Low2024,AlonsoLinaje2025}. QROM also appears in quantum solvers
for differential equations~\cite{Jennings2024,Pocrnic2025}
and linear systems~\cite{AnLin2022,Jennings2025}.
Given a table of $N$ strings $f(x)\in\{0,1\}^b$, quantum read only
memory (QROM) implements
$U_f\ket{x}\ket{y}=\ket{x}\ket{y\xor f(x)}$.
Its non-Clifford cost can be reduced by borrowing \emph{dirty qubits}:
ancillas in unknown states that must be returned unchanged, including
any entanglement with other registers.

SelectSwap uses dirty workspace to load the table in blocks of
$\lambda$ entries~\cite{Low2024,Berry2019}. With approximately
$b\lambda$ dirty qubits, its leading Toffoli count is $2N/\lambda$.
We focus on the qubit-constrained regime, where the available workspace
limits the block size enough that loading these blocks dominates the
remaining operations. The recent SelectCopy construction of
Ref.~\cite{MotlaghPocrnic2026} replaces
controlled swaps by copies and loads the output in sequential groups
of bits, or \emph{bit packets}.
Sharing each packet's unload with the next packet's load reduces the
leading count to $(1+1/b)N/\lambda$ for one-bit packets at approximately
the same dirty-qubit budget.

\begin{figure}[!t]
 \centering
 \includegraphics[width=0.97\textwidth]{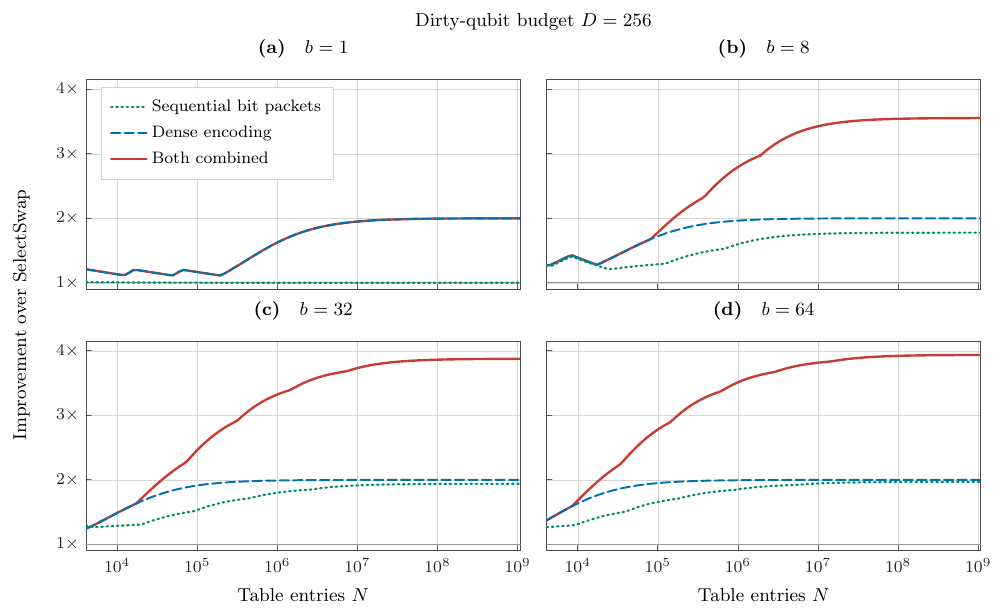}
 \caption{Toffoli reduction relative to SelectSwap with $D=256$ available
 dirty qubits. A factor of $2\times$ means half as many Toffoli gates;
 $4\times$ means one quarter as many.
 Each curve divides the independently optimized SelectSwap count
 by the optimized count of the indicated construction.
 Sequential bit packets use Ref.~\cite{MotlaghPocrnic2026}; dense encoding
 uses one full-word packet; both combined optimize the packet width as
 well as the block size. All methods have $b$ zero-initialized output
 qubits and at most $\ceil{\log_2N}$ additional clean address/helper
 qubits. Block sizes are powers of two, and all allowed integer packet
 widths are tested. The dense-encoding bounds are given in
 \cref{eq:generalbound}; the baseline bounds are taken from
 Refs.~\cite{Berry2019,MotlaghPocrnic2026}.}
 \label{fig:resources}
\end{figure}

Here we introduce \emph{dense encoding}, where we temporarily store
two bits of classical information in each dirty qubit instead of one
by utilizing both the $Z$ and $X$ bases. This enables us to load the
table in blocks of $2\lambda$ entries instead of $\lambda$ using the
same budget of $b\lambda$ dirty qubits, thereby halving the Toffoli
count of the leading load operation. Furthermore, this construction
can be combined with the idea of sequential bit packets:
\begin{equation}
 \begin{tikzpicture}[
 baseline=(current bounding box.center),
 every node/.style={align=center,inner sep=4pt},
 >={Stealth[length=4pt]},line width=0.5pt]
 \node (ss) at (0,0)
  {{\small SelectSwap}\\[5pt]$\displaystyle 2\frac{N}{\lambda}$};
 \node (seq) at (4.7,0.95)
  {{\small Sequential bit packets~\cite{MotlaghPocrnic2026}}\\[5pt]
   $\displaystyle \left(1+\frac1b\right)\frac{N}{\lambda}$};
 \node (dense) at (4.7,-0.95)
  {{\small Dense encoding (this work)}\\[5pt]
   $\displaystyle \frac{N}{\lambda}$};
 \node (both) at (10.2,0)
  {{\small Both combined (this work)}\\[5pt]
   $\displaystyle \frac12\left(1+\frac1b\right)\frac{N}{\lambda}$};
 \draw[->] (ss.east) to[out=0,in=180] (seq.west);
 \draw[->] (ss.east) to[out=0,in=180] (dense.west);
 \draw[->] (seq.east) to[out=0,in=180] (both.west);
 \draw[->] (dense.east) to[out=0,in=180] (both.west);
\end{tikzpicture}

 \label{eq:progression}
\end{equation}
The two improvements act independently: sequential packets share table
passes, whereas dense encoding increases the entries covered by each
pass. The ratio of the leading SelectSwap and combined counts is
$4b/(b+1)$, about $3.9\times$ at $b=32$, while it remains $2\times$
at $b=1$.
\Cref{fig:resources} compares the resulting improvements over SelectSwap.

\clearpage

\section{Algorithm}
\label{sec:algorithm}

\subsection{Dense encoding}
\label{sec:dense}

Let $a,z\in\{0,1\}$ be the two table bits at the same output position
of adjacent entries. Their dense encoding on one dirty qubit is the operation
$P=X^aZ^z$.
For an address bit $p$, define the selected Pauli
\begin{equation}
 B_p=\begin{cases}Z,&p=0,\\X,&p=1,\end{cases}
 \qquad f_p=(1-p)a\xor pz .
 \label{eq:selectedpauli}
\end{equation}
The read $R$ is controlled by the output qubit: it applies $B_p$ when
that qubit is in $\ket1$ and does nothing when it is in $\ket0$.
For an output basis state $\ket v$, with $v\in\{0,1\}$, its action
on the dirty qubit is therefore $R=B_p^v$.
The identities $P^\dagger ZP=(-1)^aZ$ and
$P^\dagger XP=(-1)^zX$ give
\begin{equation}
 P^\dagger RPR=(-1)^{v f_p}I.
 \label{eq:denseidentity}
\end{equation}
Thus the sequence $R,P,R,P^\dagger$ returns the dirty qubit to its
initial state and applies $Z^{f_p}$ to the output qubit.
A Hadamard before and after this sequence gives
$HZ^{f_p}H=X^{f_p}$, writing the selected bit by XOR, as in
\cref{fig:circuits}(a).

\begin{figure}[t]
 \centering
 \begin{tikzpicture}[x=1.08cm,y=0.80cm,
 every node/.style={font=\small},
 gate/.style={draw,fill=white,minimum width=0.68cm,minimum height=0.50cm},
 line width=0.45pt]
 \node[anchor=west,font=\small\bfseries] at (-0.95,4.55)
  {(a) Dense encoding of two table entries};
 \foreach \y in {3,2,1} {\draw (0,\y)--(12,\y);}
 \node[anchor=east] at (-0.20,3) {$\ket p$};
 \node[anchor=east] at (-0.20,2) {$\ket y$};
 \node[anchor=east] at (-0.20,1) {$\ket{\psi}$};
 \foreach \x in {0.6,11.3} {\node[gate] at (\x,2) {$H$};}
 \foreach \x in {2.2,6.4} {
  \draw (\x,3)--(\x,1);
  \draw[fill=white] (\x,3) circle (2.2pt);
  \foreach \y in {2,1} {\fill (\x,\y) circle (2.2pt);}
 }
 \foreach \x in {3.4,7.6} {
  \draw (\x,3)--(\x,1);
  \foreach \y in {3,2} {\fill (\x,\y) circle (2.2pt);}
  \draw[fill=white] (\x,1) circle (4pt);
  \draw (\x-0.13,1)--(\x+0.13,1);
  \draw (\x,0.825)--(\x,1.175);
 }
 \node[gate,minimum width=1.55cm] at (4.9,1) {$P=X^aZ^z$};
 \node[gate,minimum width=0.98cm] at (9.45,1) {$P^\dagger$};
 \draw[decorate,decoration={brace,amplitude=3pt}]
  (1.95,3.40)--(3.65,3.40) node[midway,above=2pt] {$R$};
 \draw[decorate,decoration={brace,amplitude=3pt}]
  (6.15,3.40)--(7.85,3.40) node[midway,above=2pt] {$R$};
 \node[anchor=west] at (12.20,3) {$\ket p$};
 \node[anchor=west] at (12.20,2) {$\ket{y\xor f_p}$};
 \node[anchor=west] at (12.20,1) {$\ket{\psi}$};

 \begin{scope}[yshift=-0.25cm]
 \node[anchor=west,font=\small\bfseries] at (-0.95,0.0)
  {(b) Full QROM with dense encoding and one packet};
 \foreach \y in {-1,-2,-3,-4} {\draw (0,\y)--(12,\y);}
 \node[anchor=east] at (-0.20,-1) {$\ket q$};
 \node[anchor=east] at (-0.20,-2) {$\ket{r,p}$};
 \node[anchor=east] at (-0.20,-3) {$\ket y$};
 \node[anchor=east] at (-0.20,-4) {$\ket{\psi}_E$};
 \foreach \x in {0.6,11.3} {
  \node[gate,minimum width=0.62cm] at (\x,-3) {$H^{\otimes b}$};
 }
 \foreach \x in {2.4,6.6} {
  \draw[fill=white] (\x-0.46,-4.35) rectangle (\x+0.46,-1.65);
  \node at (\x,-3) {$\mathcal R$};
 }
 \foreach \x/\lab in {4.55/{\mathcal P(q)},9.1/{\mathcal P(q)^\dagger}} {
  \draw[fill=white] (\x-0.68,-4.35) rectangle (\x+0.68,-0.65);
  \node at (\x,-2.5) {$\lab$};
 }
 \node[anchor=west] at (12.20,-1) {$\ket q$};
 \node[anchor=west] at (12.20,-2) {$\ket{r,p}$};
 \node[anchor=west] at (12.20,-3) {$\ket{y\xor f(x)}$};
 \node[anchor=west] at (12.20,-4) {$\ket{\psi}_E$};
 \end{scope}
\end{tikzpicture}
 \caption{Dense encoding and QROM without sequential packets.
 (a) A dirty qubit carries the Pauli choices for two table values $a,z$.
 The read $R$ applies $Z$ for $p=0$ and $X$ for $p=1$, controlled
 by the output qubit.
 (b) The full lookup uses $x=2\lambda q+2r+p$.
 $\mathcal P(q)$ loads both entries of each pair into a bank of
 $b\lambda$ dirty qubits. $\mathcal R$ selects row $r$ and the
 Pauli component determined by $p$. The final load is the true inverse
 $\mathcal P^\dagger$.}
 \label{fig:circuits}
\end{figure}

Equation~\eqref{eq:denseidentity} holds in each address sector and
therefore for coherent superpositions of $p$. Its identity action
on the dirty register also preserves arbitrary entanglement with
other registers.

To obtain full QROM, write the address as
\begin{equation}
 x=2\lambda q+2r+p,\qquad 0\leq r<\lambda,\quad p\in\{0,1\}.
 \label{eq:address}
\end{equation}
For clarity, take $\lambda$ to be a power of two and $2\lambda$ to
divide $N$; for other table sizes, pad the final block with zero labels.
Allocate $b$ dirty qubits for each of the $\lambda$ rows.
For output position $c$ and row $\ell$, define
\begin{equation}
 a_c(q,\ell)=f_c(2\lambda q+2\ell),\qquad
 z_c(q,\ell)=f_c(2\lambda q+2\ell+1).
 \label{eq:tablepairs}
\end{equation}
One high-address pass applies
$\mathcal P(q)=\prod_{c,\ell}X_{c,\ell}^{a_c(q,\ell)}
Z_{c,\ell}^{z_c(q,\ell)}$.
The table-dependent $X$ and $Z$ gates share the same high-address
indicator, so loading the second entry adds Clifford gates but no
additional Toffoli gate to that pass.

The full read $\mathcal R=\prod_c B_{p,c,r}^{v_c}$ repeats the
elementary read on each qubit in row $r$, controlled by the corresponding
output bit $v_c$. Unary iteration~\cite{Babbush2018} selects the row,
and the address bit $p$ chooses the $X$ or $Z$ read.

Applying \cref{eq:denseidentity} to every output position proves
that the circuit in \cref{fig:circuits}(b) implements the lookup
and restores all $b\lambda$ dirty qubits.
Using the Toffoli-count convention of Ref.~\cite{MotlaghPocrnic2026},
the two high passes,
two row passes, and controlled reads give the bound
\begin{equation}
 \tau_{\mathrm{dense}}(1,\lambda)
 \leq\frac{N}{\lambda}+(4b+2)\lambda-6,
 \qquad D_{\mathrm{dense}}=b\lambda.
 \label{eq:singlecost}
\end{equation}
The leading term is $N/\lambda$ for every $b\geq1$.
\Cref{fig:readhelper} shows the selected-row read used in this count.

\subsection{Dense encoding with sequential bit packets}
\label{sec:packets}

Divide the output into $\alpha$ packets of $\mu=b/\alpha$ bits.
The same $b\lambda$ dirty qubits now provide $L=\alpha\lambda$
rows, and each table-loading pass covers $2\alpha\lambda$ entries.
Here $\lambda=L/\alpha$ expresses the dirty workspace per output bit;
only $L$ must be a power of two.
As in the sequential construction of Ref.~\cite{MotlaghPocrnic2026},
the unload of packet $j-1$ and the load of packet $j$ combine into
one classically specified operation $\mathcal P_j\mathcal
P_{j-1}^\dagger$. There are therefore $\alpha+1$ high-address passes.

As in the sequential construction of Ref.~\cite{MotlaghPocrnic2026},
all initial reads can also be implemented in a single pass at the
beginning, giving only $\alpha+1$ read passes in total.

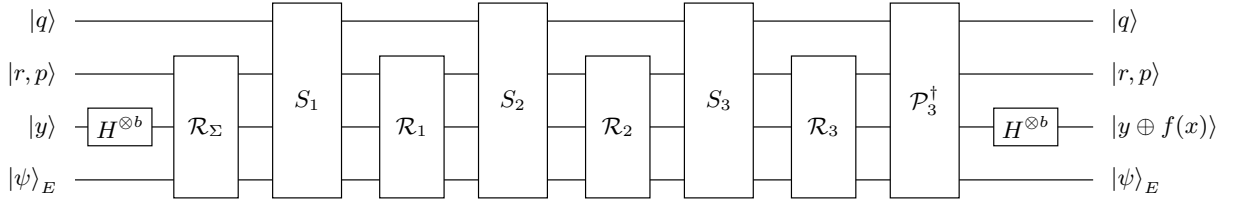
\begin{figure}[t]
 \centering
 \begin{tikzpicture}[x=0.99cm,y=0.70cm,
 every node/.style={font=\small},
 gate/.style={draw,fill=white,minimum width=0.64cm,minimum height=0.46cm},
 line width=0.45pt]
 \foreach \y in {0,-1,-2,-3} {\draw (0,\y)--(13.6,\y);}
 \node[anchor=east] at (-0.12,0) {$\ket q$};
 \node[anchor=east] at (-0.12,-1) {$\ket{r,p}$};
 \node[anchor=east] at (-0.12,-2) {$\ket y$};
 \node[anchor=east] at (-0.12,-3) {$\ket{\psi}_E$};
 \foreach \x in {0.60,12.70} {
  \node[gate] at (\x,-2) {$H^{\otimes b}$};
 }
 \foreach \x/\lab in
  {1.75/{\mathcal R_\Sigma},4.50/{\mathcal R_1},
   7.25/{\mathcal R_2},10.0/{\mathcal R_3}} {
  \draw[fill=white] (\x-0.43,-3.34) rectangle (\x+0.43,-0.66);
  \node at (\x,-2) {$\lab$};
 }
 \foreach \x/\lab in
  {3.10/{S_1},5.85/{S_2},8.60/{S_3},11.35/{\mathcal P_3^\dagger}} {
  \draw[fill=white] (\x-0.46,-3.34) rectangle (\x+0.46,0.34);
  \node at (\x,-1.5) {$\lab$};
 }
 \node[anchor=west] at (13.72,0) {$\ket q$};
 \node[anchor=west] at (13.72,-1) {$\ket{r,p}$};
 \node[anchor=west] at (13.72,-2) {$\ket{y\xor f(x)}$};
 \node[anchor=west] at (13.72,-3) {$\ket{\psi}_E$};
\end{tikzpicture}
 \caption{Dense encoding combined with three sequential bit packets.
 The initial $\mathcal R_\Sigma$ combines the packet reads using
 output XORs, formed and restored by CNOTs within this block.
 The loads are $S_1=\mathcal P_1$,
 $S_2=\mathcal P_2\mathcal P_1^\dagger$, and
 $S_3=\mathcal P_3\mathcal P_2^\dagger$.
 Each $\mathcal R_j$ reads packet $j$ from the row and Pauli component
 selected by $(r,p)$. The final $\mathcal P_3^\dagger$ restores the entire
 dirty register.}
 \label{fig:packetcircuit}
\end{figure}

The remaining steps load and read each packet, followed by one final
unload, as shown for three packets in \cref{fig:packetcircuit}.

Combining the high passes, row passes, and controlled reads yields
\begin{equation}
 \tau_{\mathrm{dense}}(\alpha,\lambda)
 \leq\left(\frac12+\frac{1}{2\alpha}\right)\frac{N}{\lambda}
   +(\alpha+1)(2b+\alpha)\lambda-3(\alpha+1),
 \qquad D_{\mathrm{dense}}=b\lambda.
 \label{eq:normalized}
\end{equation}
Here $\alpha$ divides $b$, $L=\alpha\lambda$ is a power of two,
and $2L$ divides $N$ with $2L<N$.

At $\alpha=b$, every packet contains one output bit. The full count is
\begin{equation}
 \tau_{\mathrm{dense}}(b,\lambda)
 =\frac12\left(1+\frac1b\right)\frac{N}{\lambda}
   +3b(b+1)\lambda-3(b+1).
 \label{eq:onebitpackets}
\end{equation}
Its leading term is half the sequential-packet leading cost for any
output width.
When $b=1$, the single-packet and combined constructions coincide.

For arbitrary table sizes and packet widths, \emph{integer bounds}
use whole-number counts of packets and table blocks, rounding up when
the last packet or block is incomplete.
For the bounds used in \cref{fig:resources}, take integer
$1\leq\mu\leq b$ and power-of-two $1<L<N/2$, and put
$m=\ceil{b/\mu}$ and $Q=\ceil{N/(2L)}$. Then
\begin{equation}
 \tau_{\mathrm{dense}}\leq(m+1)(Q+L-3)+2L(b+\mu),
 \qquad D_{\mathrm{dense}}=\mu L.
 \label{eq:generalbound}
\end{equation}
Incomplete packets and table blocks receive zero labels, with the
promised address range $x<N$.
Clean address and gate helpers are reused between passes; their number
is at most $\max\{\ceil{\log_2Q},\log_2(2L)\}$.

\vspace{-2\baselineskip}
\section{Conclusion}

Dense encoding uses the two Pauli components of each dirty qubit to
load bits from two table entries. It reduces the
leading SelectSwap Toffoli count to $N/\lambda$ without an output-width
restriction. Combining it with sequential bit packets gives
$\tfrac12(1+1/b)N/\lambda$: a second halving of the sequential
construction's leading cost and an improvement of approximately
$4\times$ over SelectSwap for large output widths.
Future work should extend the comparison to Clifford scheduling,
measurement latency, and physical layout to determine how these logical
savings translate to particular fault-tolerant implementations.

\paragraph*{Acknowledgments.}
Generative AI tools assisted in conceptualizing the dense-encoding
approach, developing and checking the construction, and preparing this
manuscript. The author takes full accountability for the correctness of
the result. The author additionally thanks Matt Pocrnic and Juan Miguel
Arrazola for their independent verification of the construction.

\clearpage
\bibliography{references}
\appendix
\crefalias{section}{appendix}
\setlength{\parskip}{0.5\baselineskip}

\section{Selected-row read circuit}
\label{app:readhelper}

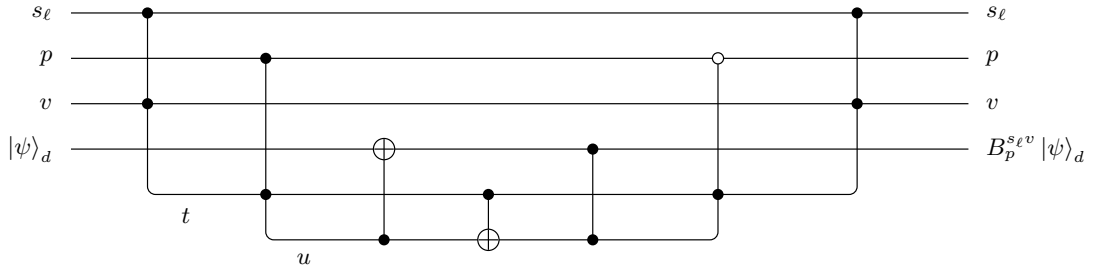
\begin{figure}[!h]
 \centering
 \begin{tikzpicture}[x=0.92cm,y=0.60cm,
 every node/.style={font=\small},
 line width=0.45pt]
 \foreach \y in {0,-1,-2,-3} {\draw (0,\y)--(12.9,\y);}
 \node[anchor=east] at (-0.14,0) {$s_\ell$};
 \node[anchor=east] at (-0.14,-1) {$p$};
 \node[anchor=east] at (-0.14,-2) {$v$};
 \node[anchor=east] at (-0.14,-3) {$\ket\psi_d$};

 \draw[rounded corners=3pt] (1.1,0)--(1.1,-4)--(11.3,-4)--(11.3,0);
 \foreach \y in {0,-2} {\fill (1.1,\y) circle (2.1pt);}
 \node[anchor=north] at (1.65,-4.12) {$t$};
 \draw[rounded corners=3pt] (2.8,-1)--(2.8,-5)--(9.3,-5)--(9.3,-1);
 \foreach \y in {-1,-4} {\fill (2.8,\y) circle (2.1pt);}
 \node[anchor=north] at (3.35,-5.12) {$u$};

 \draw (4.5,-3)--(4.5,-5);
 \fill (4.5,-5) circle (2.1pt);
 \draw[fill=white] (4.5,-3) circle (4pt);
 \draw (4.36,-3)--(4.64,-3);
 \draw (4.5,-3.22)--(4.5,-2.78);

 \draw (6.0,-4)--(6.0,-5);
 \fill (6.0,-4) circle (2.1pt);
 \draw[fill=white] (6.0,-5) circle (4pt);
 \draw (5.86,-5)--(6.14,-5);
 \draw (6.0,-5.22)--(6.0,-4.78);

 \draw (7.5,-3)--(7.5,-5);
 \foreach \y in {-3,-5} {\fill (7.5,\y) circle (2.1pt);}

 \draw[fill=white] (9.3,-1) circle (2.1pt);
 \fill (9.3,-4) circle (2.1pt);
 \foreach \y in {0,-2} {\fill (11.3,\y) circle (2.1pt);}

 \node[anchor=west] at (13.03,0) {$s_\ell$};
 \node[anchor=west] at (13.03,-1) {$p$};
 \node[anchor=west] at (13.03,-2) {$v$};
 \node[anchor=west] at (13.03,-3) {$B_p^{s_\ell v}\ket\psi_d$};
\end{tikzpicture}
 \caption{Selected-row Pauli read.
 For row indicator $s_\ell=[r=\ell]$ and output control $v$, the helpers
 store $t=s_\ell v$ and $u=pt$. The middle CNOT changes $u$ to
 $(1-p)t$, selecting $X$ or $Z$ on the dirty target according to $p$.
 Elbows denote temporary ANDs and their measurement-based uncomputations
 \cite{Gidney2018}. Each AND counts as one Toffoli unit; its uncomputation
 uses only Clifford operations.}
 \label{fig:readhelper}
\end{figure}

\end{document}